# Space-resolved characterization of high frequency atmospheric-pressure plasma in nitrogen applying optical emission spectroscopy and numerical simulation

**Priyadarshini Rajasekaran, Cornelia Ruhrmann, Nikita Bibinov, Peter Awakowicz**

Institute for Electrical Engineering and Plasma Technology, Ruhr-Universität Bochum, Universitätstr. 150, 44801 Bochum, Germany

Email: *rajasekaran@aept.rub.de, ruhrmann@aept.rub.de, nikita.bibinov@rub.de, awakowicz@aept.rub.de*

**Abstract.** *Averaged* plasma parameters such as electron distribution function and electron density are determined by characterization of high frequency (2.4 GHz) nitrogen-plasma using both experimental methods, namely optical emission spectroscopy (OES) and microphotography, and numerical simulation. Both direct and stepwise electron-impact excitation of nitrogen emissions are considered. The determination of *space-resolved* electron distribution function, electron density, rate constant for electron-impact dissociation of nitrogen molecule and the production of nitrogen atoms, applying the same methods, is discussed. Spatial distribution of intensities of neutral nitrogen molecule and nitrogen molecular ion from the microplasma is imaged by a CCD camera. The CCD images are calibrated using the corresponding emissions measured by absolutely-calibrated OES, and are then subjected to inverse Abel transformation to determine space-resolved intensities and other parameters. The *space-resolved* parameters are compared, respectively, with the *averaged* parameters, and an agreement between them is established.

PACS: 52.50 Dg, 52.70.Kz

## 1. Introduction

Atmospheric-pressure plasma devices are widely investigated for numerous diversified applications. For effective optimization of plasma sources, determination of plasma parameters such as electron density and electron distribution function is important. To this end, we have characterized different plasma sources devised for skin therapy [1, 2, 3, 4] and for deposition of thin films on the inner walls of small-diameter tubes [5]. Experimental methods namely optical emission spectroscopy (OES), current voltage measurements and microphotography, and numerical simulation are employed for plasma characterization. Characterization of plasmas applying OES supports simulation of chemical kinetics, and thereby the determination of fluxes of chemically-active species produced by plasma-chemical reactions.

In the OES-based characterization method, the two nitrogen lines namely $N_2$(C-B,0-0) at 337.1 nm and $N_2^+$(B-X,0-0) at 391.4 nm are used. Averaged gas temperature in the active plasma volume is determined from the rotational distribution of $N_2$(C-B,0-0). The ratio of measured intensities of these two nitrogen lines are used for determination of the averaged electric field in the plasma [6], and the averaged electron density in the active plasma volume is calculated from the measured absolute intensity of $N_2$(C-B,0-0). *Averaged* parameters can produce erroneous results (such as incorrect excitation rates). For instance [7], at specific conditions, the rate constant for dissociation of nitrogen molecule determined using averaged parameters is, by a factor of 2, larger than the same rate constant calculated using space-resolved plasma parameters.

On the other hand, for OES characterization of high frequency (HF) discharge (2.4 GHz) in nitrogen and nitrogen-oxygen mixture [4] only direct excitation by electron-impact of ground state $N_2(^1\Sigma_g^+)$ was considered. The influence of stepwise excitation of nitrogen emission on the reliability of determined plasma parameters was not discussed.

Hence, in this paper, first we analyse the reliability of excitation model of nitrogen emission for characterization of HF discharge in nitrogen at atmospheric pressure. Secondly, we compare the *averaged* and *space-resolved* plasma parameters. OES characterization with space resolution is realized applying CCD camera with band filters which is calibrated using the spectrometer. Numerical simulation and OES measurements supplement each other for the determination of plasma parameters.

## 2. Experimental arrangement and plasma diagnostics

### 2.1. Microwave plasma source

The atmospheric-pressure microwave plasma source [4] is a miniature device comprising of a cylindrical hollow copper resonator and a power generator. Figure 1 shows the schematic illustration of the microwave plasma source. The resonator is 12 mm long with an outer diameter of 8 mm and wall thickness of 1 mm. The resonator has a nozzle which is 0.6 mm in diameter. A thin copper wire of diameter 0.68 mm is housed inside the resonator along the longitudinal axis of the cylinder.

The "on-board" power generator of the plasma source is a gallium nitride high electron mobility transistor (GaN-HEMT) powered by 24 V DC supply. The frequency of the oscillations produced is about 2.4 GHz. The secondary winding of the transformer is connected to the copper wire of the resonator. Plasma is ignited on the tip of the copper wire due to the high voltage generated across the secondary winding of the transformer. Gas flow through the resonator is regulated using suitable flow controllers. Depending on the gas flow rate, the active plasma volume ignited on the tip of the copper wire is blown out of the resonator through the nozzle as a turbulent effluent. For the experiments reported here, nitrogen is used as the working gas flowing at a rate of 600 sccm, and diagnostics is focussed on the active plasma volume observed through the quartz window. The wall of the resonator is provided with a small window through which the plasma ignited on the tip of the copper wire is visible. To protect the active plasma volume from the disturbances caused by ambient air and to facilitate optical measurements, a quartz plate is fixed to the window as shown in figure 1.

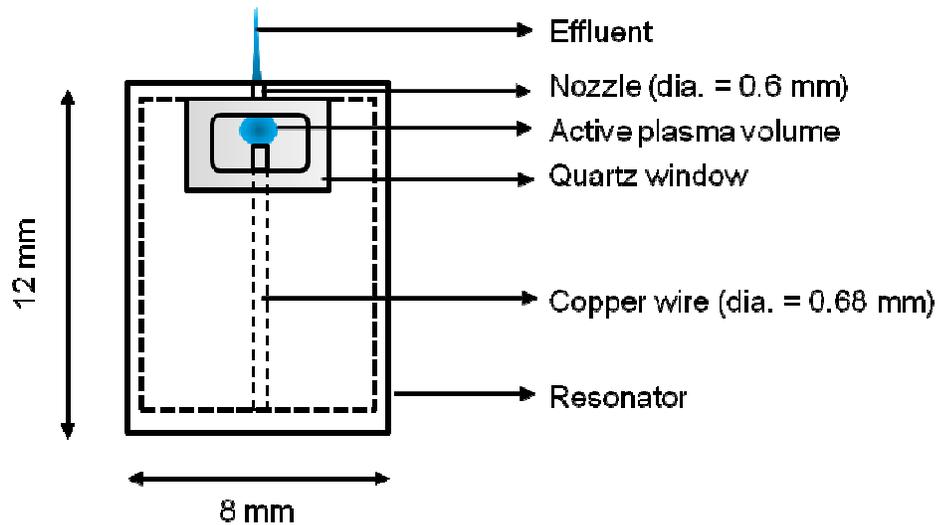

**Figure 1**. Schematic illustration of the microwave plasma source. The active plasma volume on the tip of the copper wire and the effluent exiting the nozzle of the resonator are shown.

The microwave plasma source has been profoundly investigated and the determination of plasma parameters such as electron distribution function and electron density by plasma characterization is already reported [4]. The plasma characterization method is briefly discussed in Section 3 of this paper. So far, nitrogen emissions are used for plasma characterization and are measured from the plasma volume using optical emission spectroscopy without space and time resolution. Hence, the determined plasma parameters are also averaged over space and time. However, the plasma parameters of atmospheric-pressure plasma sources strongly depend on space which is a crucial point in its applications. Therefore, space resolved measurement of the most important parameters is absolutely essential. If the nitrogen emissions are measured with space resolution, then space-resolved plasma characterization is possible for the determination of plasma parameters that are also spatially resolved.

### 2.2. Optical diagnostics

Optical emission spectroscopy (OES) of the microwave plasma is performed using an Echelle spectrometer (ESA 3000, LLA Instruments, Germany) which is relatively and absolutely calibrated using a tungsten ribbon lamp and the known molecular emission of nitrogen and nitric oxide [8]. The spectrometer has a spectral resolution of R=13333 which corresponds to the FWHM of the apparatus-function of 0.015-0.06 nm in the range of 200-800 nm. The optic fibre of the spectrometer is positioned close to the quartz window of the resonator to measure emissions from the microplasma ignited on the tip of the copper wire. The acceptance angle of the optical fibre (about 3.5°) is big enough to assume the active plasma volume on the tip of the copper wire of HF resonator as a point-like light source.

### 2.3. Microphotography

Microphotography is used to determine the volume of the active plasma region by imaging. A high-speed sensitive CCD camera (PCO Sensicam qe, PCO, Germany) is used. The exposure time of the camera can be varied from a minimum of 500 ns to about 1 hour. A suitable zoom objective (Zoom 6000®, Navitar Inc., USA) is fitted to the CCD camera and adjusted to image the microplasma as shown in figure 2. From the image, the averaged volume of the active plasma is determined considering cylinders of different diameters and heights placed one over the other. The averaged plasma volume is determined to be $5.44 \cdot 10^{-11}$ m$^3$. The edges of the plasma volume are determined from the intensity distribution, observed by microphotography, using the region with intensity distinctly higher than the background.

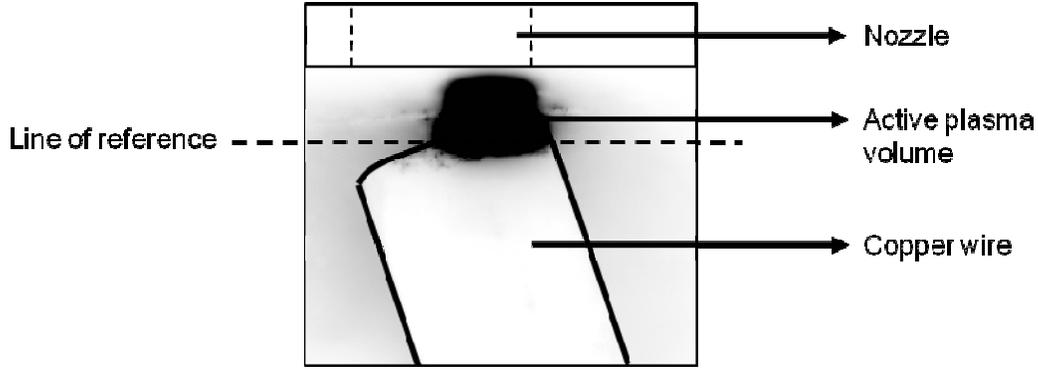

**Figure 2.** Inverted CCD-image of the active plasma volume ignited on the tip of the copper wire as seen through the quartz window by a CCD camera.

### 3. Plasma characterization

The determination of plasma parameters such as electron distribution function and electron density, and gas temperature in the active plasma volume is termed as "Plasma characterization". The plasma characterization-method applying OES and numerical simulation for nitrogen-containing plasmas, and the plasma-chemical model used for characterization have been presented already [1,9].

The averaged gas temperature in the active plasma volume is determined using OES and numerical simulation applying emission spectrum $N_2$(C-B,0-0) at 337.1 nm in assumption that rotational and translational degrees of freedom of diatomic molecules have equal temperature at atmospheric pressure conditions. The measured spectrum is compared with spectrum that is simulated for different rotational temperatures, and from the best fit, the averaged gas temperature in the active plasma volume is determined as $1000 \pm 50$ K.

Nitrogen molecular emission $N_2$(C-B) and $N_2^+$(B-X) can be excited directly (1,3) by electron impact of $N_2(X^1\Sigma_g^+)$, and step-wise through nitrogen neutral metastable (2) and ground state of nitrogen molecular ion (4)

$$N_2(X^1\Sigma_g^+) + e \rightarrow N_2(C^3\Pi_u) + e \tag{1}$$

$$N_2(A^3\Sigma_u^+) + e \rightarrow N_2(C^3\Pi_u) + e \tag{2}$$

$$N_2(X^1\Sigma_g^+) + e \rightarrow N_2^+(B^2\Sigma_u^+) + 2e \tag{3}$$

$$N_2^+(X^2\Sigma_g^+) + e \rightarrow N_2^+(B^2\Sigma_u^+) + e \tag{4}$$

Cross sections of electron impact excitations (1, 3, 4) are well known [10,11]. We do not find in the literature the cross section of excitation of $N_2$(C-B) emission by electron-impact of nitrogen metastables (2). Hence, to estimate the rate of this excitation process, we apply the cross section of electron-impact excitation of process (5)

$$N_2(X^1\Sigma_g^+) + e \rightarrow N_2(a^1\Pi_g) + e, \tag{5}$$

where multiplicity of nitrogen molecule, angular momentum of electrons and symmetry are changed similar to excitation process (2). In this condition, cross section of process (5) is shifted, in electron kinetic-energy scale, to the threshold corresponding to the energy of transition $N_2$(A-C). Because of high rate constant of pooling reaction (6) and quenching during collisions with nitrogen atoms (7), steady state density of nitrogen metastables is more than five orders of magnitude lower than density of nitrogen in ground state, and hence the probability of step-wise excitation of $N_2$(C-B) emission is negligible at our experimental conditions.

$$N_2(A^3\Sigma_u^+) + N_2(A^3\Sigma_u^+) \rightarrow N_2(B,C) + N_2(X^1\Sigma_g^+) \tag{6}$$

$$N_2(A^3\Sigma_u^+) + N \rightarrow N_2(X^1\Sigma_g^+) + N \tag{7}$$

Plasma parameters namely, electron velocity distribution function ($f_v$(E) in eV$^{-3/2}$) and electron density ($n_e$ in m$^{-3}$), are determined by solving (8) and (9).

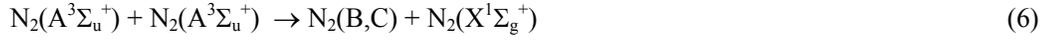

$$\frac{I_{N_2^+(B-X,0-0)}}{I_{N_2(C-B,0-0)}} = \frac{Q_{N_2^+(B)} \cdot \left(N_{N_2} \cdot k_{N_2^+(B)} + N_{N_2^+(X)} \cdot k_{N_2^+(B)}^{N_2^+}\right) \cdot n_e}{Q_{N_2(C)} \cdot N_{N_2} \cdot k_{N_2(C)} \cdot n_e} \tag{8}$$

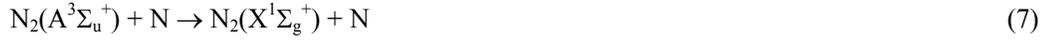

$$I_{N_2(C-B,0-0)} = Q_{N_2(C)} \cdot N_{N_2} \cdot k_{N_2(C)} \cdot V_p \cdot n_e, \tag{9}$$

where, $Q_{N_2^+(B)} = \dfrac{A'}{A' + k_{q_2}^{N_2^+(B)} \cdot N_{N_2}}$ and $Q_{N_2(C)} = \dfrac{A}{A + k_{q_1}^{N_2(C)} \cdot N_{N_2}}$, A' and A are the corresponding Einstein coefficients [12], $k_{q_2}^{N_2^+(B)}$ and $k_{q_1}^{N_2(C)}$ (in m$^3$·s$^{-1}$) are the rate constants for quenching of $N_2^+$(B) and $N_2$(C) excited states during collision with $N_2$ [8], $N_{N_2}$ (in m$^{-3}$) is the density of $N_2$ at determined gas temperature. In (8), $k_{N_2^+(B)}$ and $k_{N_2(C)}$ (in m$^3$·s$^{-1}$) are the rate constants, respectively, for electron impact excitation of $N_2$(C-B) and $N_2^+$(B-X) emission from $N_2$(X), $k_{N_2^+(B)}^{N_2^+}$ is the rate constant for electron impact excitation of $N_2^+$(B-X) emission from $N_2^+$(X), $N_{N_2^+(X)}$ (in m$^{-3}$) is the density of nitrogen ions in ground state, $n_e$ (in m$^{-3}$) is the electron density that cancels out, and $V_p$ (in m$^3$) is the plasma volume.

In assumption that the plasma is quasi-neutral and density of $N_2^+$(X) is approximately equal to the electron density, (8) and (9) can be transformed to (10) and (11).

$$n_e = \frac{\dfrac{I_{N_2^+(B-X,0-0)}}{I_{N_2(C-B,0-0)}} \cdot N_{N_2} \cdot \left(Q_{N_2(C)} \cdot k_{N_2(C)} - Q_{N_2^+(B)} \cdot k_{N_2^+(B)}\right)}{Q_{N_2^+(B)} \cdot k_{N_2^+(B)}^{N_2^+}} \tag{10}$$

$$n_e = \frac{I_{N_2(C-B,0-0)}}{Q_{N_2(C)} \cdot N_{N_2} \cdot k_{N_2(C)} \cdot V_p} \tag{11}$$

The rate constants of electron impact excitations are functions of electric field. To calculate these constants, the Boltzmann equation is solved numerically in "local approximation" at electric field oscillation with a frequency of 2.4 GHz and atmospheric-pressure condition in nitrogen for different values of electric field $E_0$ since the local field strength in the plasma is not known. $k_{N_2^+(B)}$, $k_{N_2(C)}$ and $k_{N_2^+(B)}^{N_2^+}$ are dependent on EVDF and on the corresponding cross-section ($\sigma$ in $m^2$). These constants are calculated using (12):

$$k = 4\pi\sqrt{2}\int_0^\infty f_v(E) \sqrt{\frac{2e}{m}} E \cdot \sigma \ (E) \, dE \tag{12}$$

where, m (in kg) is the mass of electron, e (in Coulomb) is the elementary charge of electron, and E (in eV) is the kinetic energy of electrons. $f_v(E)$ is normalized to fulfill (13):

$$4\pi\sqrt{2}\int_0^\infty f_v(E) \sqrt{E} \ dE = 1 \tag{13}$$

$k_{N_2^+(B)}$, $k_{N_2(C)}$ and $k_{N_2^+(B)}^{N_2^+}$ are calculated for variable $E_0$, and (10) and (11) are presented graphically in the plot $n_e$ vs. high frequency electric field $E_0$ (figure 3).

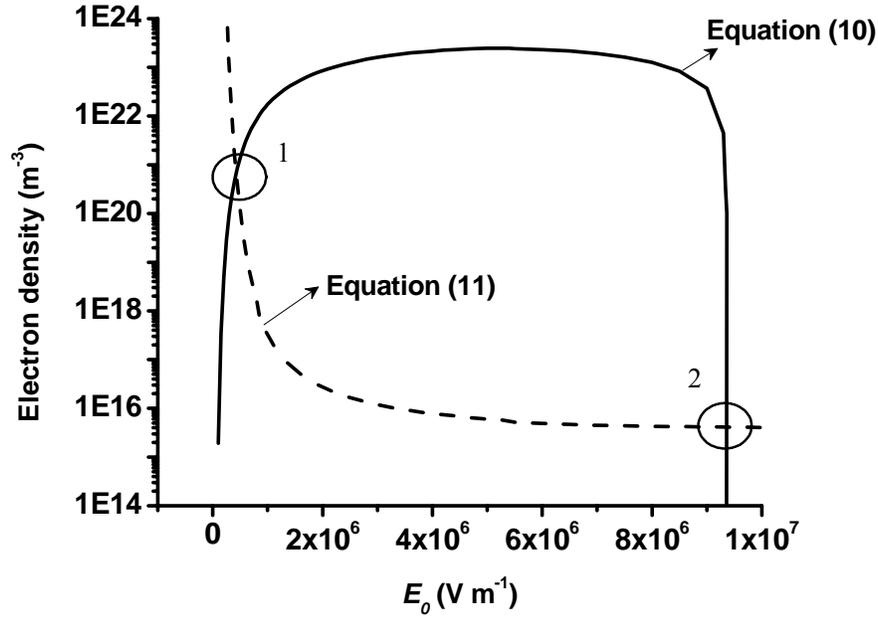

**Fig 3.** Electron density ($n_e$ in $m^{-3}$) calculated using (10) and (11) for different $E_0$ fields (in V·$m^{-1}$). The intersecting points (1) and (2) that correspond to two solutions of the equation system (10,11) are marked with circles.

From figure (3), it is clear that the equation system (10,11) has two solutions, one at low electric field ($E_0 = 4.2 \cdot 10^5$ V·$m^{-1}$) and the other at high electric field ($E_0 = 9.4 \cdot 10^6$ V·$m^{-1}$). To find reliable plasma parameters additional information is needed. For this purpose, we apply rotational distributions in emission bands of neutral and ionized nitrogen molecules. As was mentioned above, rotational temperature in emission spectrum of $N_2$(C-B) is equal to gas temperature because of high efficiency of rotational relaxation in neural nitrogen molecules at atmospheric pressure conditions. At the same time, rotational relaxation in excited state $N_2^+$(B) does not occur because almost each collision with ambient neutral nitrogen molecules results in quenching of excitation. Since, the nuclear momentum is changed slightly during collisions with electrons and $N_2^+$(B) state cannot relax in our experimental conditions, the rotational distribution in the $N_2^+$(B-X) emission spectrum corresponds to the rotational distribution of species before electron-impact excitation. The rotational temperature determined in emission spectrum of ionized nitrogen molecule $N_2^+$(B-X) at our experimental conditions ($T_r = 1830 \pm 30$ K) is much higher than gas temperature measured applying emission band of neutral nitrogen molecule $N_2$(C-B) ($T_g = 1000 \pm 50$ K). The nitrogen molecular ions in ground state $N_2^+$(X) are accelerated in the electric field and have higher kinetic energy than the neutrals. Because of numerous collisions (mainly with neutrals) at atmospheric pressure conditions, rotational and translational degrees of freedom of nitrogen molecular ions have equal temperature that is higher than the gas temperature [13]. On the basis of the difference in rotational temperature determined in nitrogen emissions, we conclude that at our experimental conditions $N_2^+$(B-X) emission is excited mainly by step wise excitation via $N_2^+(X)$ state. However, according to our calculations, step wise excitation of $N_2^+(B-X)$ emission at high electric field ($E_0 = 9.4 \cdot 10^6$ V·m$^{-1}$) is negligible. Therefore, solution of the equation system (10,11) at low electric field and high electron density (crossing point 1 in figure 3) is more reliable at HF plasma source conditions. The determined plasma parameters ($E_0 = 4.2 \cdot 10^5$ V·m$^{-1}$ and $n_e = 7.0 \cdot 10^{20}$ m$^{-3}$) in the HF plasma source differ strongly from those determined applying OES [4] in assumption of direct electron-impact excitation, where analysis of the excitation model was not performed.

## 4. Space-resolved plasma characterization: Determination of space-resolved plasma parameters

### 4.1. Concept of space-resolved plasma characterization

If the photoemissions from the plasma volume will be measured with space resolution, then using the above discussed characterization method, space-resolved plasma parameters can be determined. Using the Echelle spectrometer, emissions can be measured with a space resolution of about 1 mm provided the optical fibre is fitted with a diaphragm. But to measure space-resolved intensities from small plasma volumes, such as the active plasma volume of the microplasma source, a method with much higher space resolution becomes necessary for measuring emissions, which then becomes useful for the determination of plasma parameters that are spatially resolved.

Up to now, we have used a sensitive CCD camera only for the determination of volume of the active plasma region through microphotography. For the first time, we present here the space-resolved plasma characterization with the help of the same CCD camera. While imaging using the camera and a suitable interference band-pass filter, the pixels in the CCD of the camera are irradiated by the emissions from the active plasma volume. As a result, each pixel in the image contains information on the degree or level of irradiation caused by the incoming nitrogen emissions. The intensity contained in each pixel of the image is calibrated in absolute units (photons·s$^{-1}$), using nitrogen emissions measured using the calibrated spectrometer, which serves useful for the determination of plasma parameters using the characterization method discussed in Section 3.

In principle, optical imaging is achieved using the zoom objective fitted to the camera. In an ideal case, each line-of-sight in the plasma volume imaged onto the CCD-chip is a very narrow line. In reality, each line has the form of two cones aligned along this line with their apexes connected with each other. At this connecting point falls the focal point where the best spatial resolution is achieved. On the other hand, the worst resolution arises beyond the radiating zone on the borders of the plasma volume. In order to minimize these uncertainties, a long focal length is used which in our case is combined with a small plasma volume.

*4.2. Selection of interference band-pass filters for imaging*

To determine space-resolved emissions from $N_2(C,0)$ and $N_2^+(B,0)$, band pass filters $360 \pm 5$ nm and $390 \pm 5$ nm (Knight Optical UK Limited) are used. Emission bands namely $N_2(C-B,0-1)$ at 357.6 nm and $N_2^+(B-X,0-0)$ at 391.4 nm are selected. These wavelengths are chosen according to the intensity distribution in emission spectrum of nitrogen, and sensitivity of CCD camera. For the purpose of plasma characterization, absolute intensities of emission bands of $N_2(C-B,0-0)$ at 337.1 nm and $N_2^+(B-X,0-0)$ are required. The CCD images are absolutely calibrated applying these bands measured by the Echelle spectrometer. Relation between intensities of $N_2(C-B,0-0)$ and $N_2(C-B,0-1)$ bands correspond to the Franck-Condon factors of these intensities which are independent from plasma conditions and probabilities of excitation and de-excitation processes for $N_2(C,0)$ level. This relation is included in the calibration factor.

The interference filter $390 \pm 5$ nm has too broad transmission band for precise measurement of $N_2^+(B-X)$ emission. Three bands of second positive system of nitrogen, namely $N_2(C-B,2-5)$ at 394.3 nm, $N_2(C-B,3-6)$ at 389.4 nm and $N_2(C-B,4-7)$ at 385.7 nm, are also observed by the CCD camera when using this band pass filter. According to averaged measurement using the spectrometer, intensity of $N_2^+(B-X,0-0)$ amounts to 65% of total intensity measured using this filter. Systematic error in the determination of plasma parameters caused by this inaccuracy amounts to 2.5% for determination of electric field and 25% for determination of electron density. In future, band pass filters with narrow transmission band will be applied for space resolved characterization of nitrogen plasma.

*4.3. Data acquisition and processing for plasma characterization with space-resolution*

The interference filters are used one at a time for imaging the nitrogen photoemissions individually. The CCD images thus obtained with the respective filter are exported as ASCII data using the camera-control software (Camware). The size of each pixel in the image is determined as 1.2 μm from the image. The exported data will be a matrix and the number value in each cell corresponds to the intensity of the pixel in the image. This matrix gives information on the spatial distribution of CCD irradiation with a space resolution of about 1.2 μm which corresponds to the sides of a square-shaped pixel with the length equal to the broadness of the radiative volume which is in turn equal to the broadness of the region in the image where intensity is higher than the background value. The irradiation of each pixel is absolutely calibrated to obtain space-resolved absolute intensities of $N_2(C-B,0-0)$ and $N_2^+(B-X,0-0)$.

*4.4. Absolute calibration of the images using nitrogen emissions measured applying OES*

In order to calibrate the CCD-images made using the filters, emissions from the entire active plasma volume are measured without time and space resolution using OES. The measured intensities are corrected taking into account the spectral efficiency ($\varepsilon(\lambda)$) of the Echelle spectrometer and the geometric factor (G), which result in the *actual intensities* from the active plasma volume expressed in photons·s$^{-1}$. The geometric factor G is the ratio between the circular area of the entrance hole of optic fiber to the surface area of the sphere with diameter d (the distance between the optic fibre and the active plasma volume). The emission bands of $N_2$(C-B,0-0) at 337.1 nm and $N_2^+$(B-X, 0-0) at 391.4 nm are integrated in the corresponding spectral ranges.

On the other hand, images of the active plasma volume are made using suitable filters. Each pixel in the image is characterized with a grayscale value. In principle, the summation of these grayscale values (after appropriate background correction) gives the *total intensity* imaged by the CCD camera. This *total intensity* should account for the *actual intensity* determined using OES (as shown in the previous step). Hence, to calibrate the grayscale values in the image in terms of photons·s$^{-1}$, the following is performed. The *actual intensity I* (determined using OES) is divided by the *total intensity* $\sum_j I(x_j, y_j)$ (imaged by the CCD camera) for each of the nitrogen emissions, and the resulting factor (*F*), as shown in (14).

$$\frac{I \text{ (from Echelle ESA 3000)}}{\sum_j I(x_j, y_j) \text{ (from CCD image with filter)}} = F \tag{14}$$

The factor F calculated, individually, for $N_2^+$(B-X, 0-0) and $N_2$(C-B,0-0) emissions is used to correct the grayscale value in each pixel in the respective image made using the filters by multiplying the corresponding factor to the grayscale value of each pixel in the image. This results in two images which correspondingly contain the absolute intensity of $N_2^+$(B-X, 0-0) and $N_2$(C-B,0-0) in each pixel.

*4.5. Inverse Abel transformation*

The absolute intensity of the 2-D image made using microphotography is spatially-resolved only in two directions namely, the *x* and *y* directions. The absolute intensity in each pixel corresponds to the sum of individual intensities in pixels that are aligned in *z*-direction which is the orthogonal-projection of each pixel in the image. We assume a cylindrical symmetry of the plasma volume and eliminate deviation of radial intensity distribution in 2-D image from symmetric one. The rotationally symmetric 3-D radial distribution I(*r*) is reconstructed then by applying a Fourier analysis, explained in detail in [14].

*4.6. Plasma characterization with space resolution*

To determine electric field with space resolution in HF discharge in nitrogen applying space distribution of nitrogen emission, we assume direct electron impact excitation of $N_2$(C-B) emission and step-wise excitation of $N_2^+$(B-X) emission via $N_2^+$(X) state. This assumption arises from the averaged OES plasma characterization. In the frames of this excitation model, (8) and (9) can be transformed into (15):

$$\frac{I_{N_2^+(B-X,0-0)}}{I_{N_2(C-B,0-0)}^2} = \frac{Q_{N_2^+(B)} \cdot k_{N_2^+(B)}^{N_2^+}}{Q_{N_2(C)}^2 \cdot N_{N_2}^2 \cdot k_{N_2(C)}^2} \tag{15}$$

Therefore, the ratio of intensities of nitrogen emissions is a function of electric field and is calculated using (12). Applying this function and intensities of measured nitrogen emissions, we determine electric field with space resolution. Electron density is determined using (11), where $V_p$= $1.73 \cdot 10^{-18}$ m$^3$ (pixel volume).

For plasma characterisation with space resolution, we use averaged value of gas temperature determined applying OES. Electric field and electron density both depend on nitrogen density (11,15), and therefore depend on gas temperature because pressure in our experiment is constant. Applying only OES diagnostic, whose space resolution is not better than 1 mm, the space distribution of the gas temperature cannot be determined. Probably combination of numerical simulation and OES measurement can provide information necessary for space resolved plasma characterization. This will be attempted in the future.

Rate constants for electron-impact physical processes such as dissociation of nitrogen molecules and excitation of nitrogen metastables are useful for the simulation of chemical kinetics. Since the rate constants depend on $f_v(E)$, they determined using the space-resolved $f_v(E)$ values (12). Finally, space-resolved electron density in the active plasma volume is calculated using (11). The production of nitrogen atoms (in m$^{-3} \cdot$s$^{-1}$), due to electron-impact dissociation of nitrogen molecule, is determined using the corresponding rate constants, the electron density and the density of nitrogen molecule at averaged gas temperature.

## 5. Results and discussion

### 5.1 Space-resolved plasma parameters

Photoemissions are imaged by the high-speed sensitive camera employing suitable filters for the second positive system and the first negative system of nitrogen. The CCD images are calibrated (in absolute units) using nitrogen photoemission measured by OES (figure 4). Abel- transformation is performed and the resulting 3-D images are shown in figure 5.

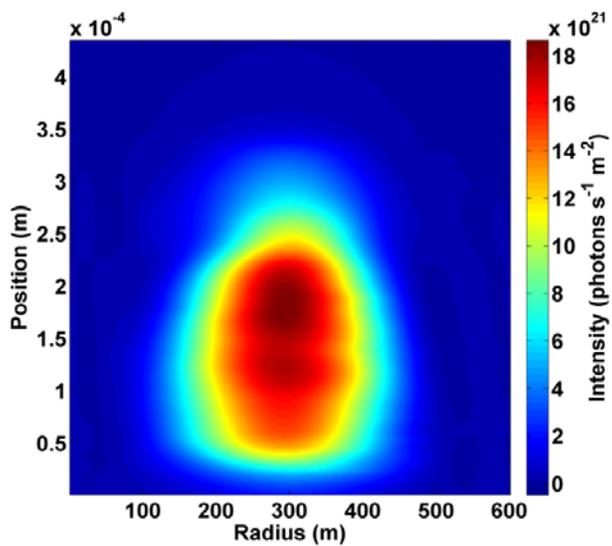

**(a)**

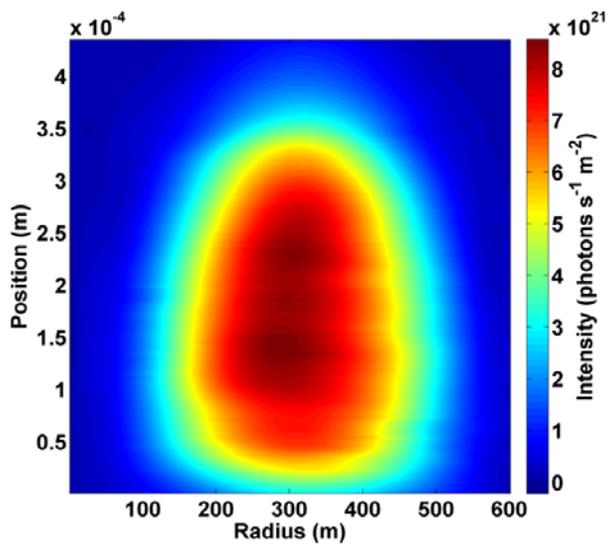

**(b)**

**Figure 4.** Intensity of (a) $N_2^+$(B-X, 0-0) and (b) $N_2$(C-B,0-0) in absolute units after absolute calibration of the CCD images using emissions measured by OES.

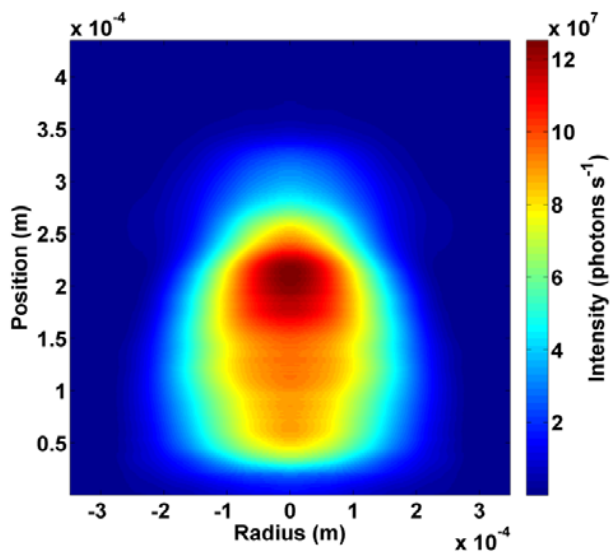

**(a)**

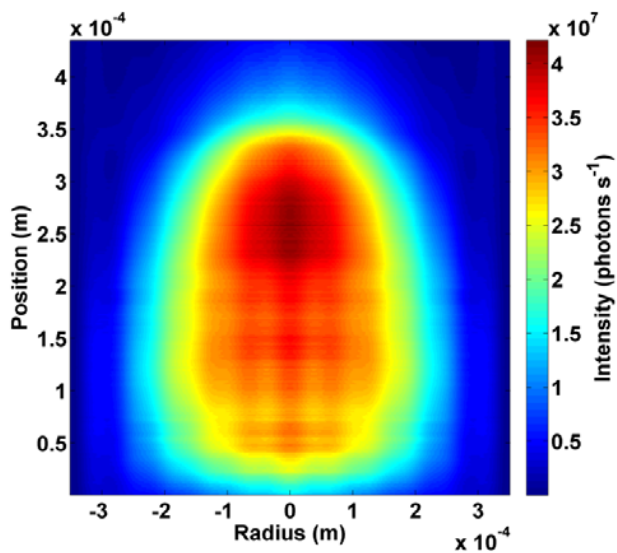

**(b)**

**Figure 5.** Absolute intensity of (a) $N_2^+$(B-X, 0-0) and (b) $N_2$(C-B,0-0) after inverse Abel transformation.

The electric field in each pixel volume is determined by evaluating the ratio of experimentally-measured intensities of $N_2^+$(B-X) to $N_2$(C-B), calculated quenching factors, simulated rate constants for electron impact excitation using (15). The ratio of rate constants for $N_2^+$(B-X) to that of $N_2$(C-B) excitation is simulated for variable HF electric field, and the polynomial relation between the ratio of simulated rate constants and local HF electric field is determined. Applying this relation, the HF electric field in each pixel volume is determined.

As a result, the electric field is determined with a space resolution of 1.2 μm. In figure 6 are shown the results of the electric field determined at different distances from the tip of the copper wire along the radius of the microplasma volume. Accordingly, only few distances are selected which correspond to 50, 150, 250 and 350 μm from the tip of the copper wire. In addition, the electric field determined using averaged OES (figure 1) $E_o = 4.2 \cdot 10^5$ V·m$^{-1}$ is also presented in figure 6 for comparison with the space-resolved electric field. It can be seen that the electric field increases in the edges of the active plasma volume. Similar results are also shown through simulation of electric fields in a microwave plasma in $N_2/O_2$ mixture [15].

The space-resolved electric field values are slightly lower than that determined applying averaged OES. The probable reason for this difference could be the influence of variations caused by gas temperature. According to (15), electric field determined using measured ratio of nitrogen emissions depends on square of nitrogen density that is function of gas temperature at constant pressure conditions. Averaged gas temperature determined applying emission spectroscopy without space resolution is used in our calculations. In this case, the electric field determined in the regions where gas temperature is higher than the "averaged" gas temperature is under-estimated, and is also overestimated in the regions which actually have lower (than "averaged") gas temperature. It is clear that the gas temperature is not constant in microplasma volume, and because of cooling process the maximum value is in the middle of active plasma volume. This could intensify the electric field on the edges of the microplasma (R > 250 μm). Therefore, space-resolved gas temperature is required for reliable determination of electric field applying space-resolved OES. As was discussed above, combination of spectroscopic measurements and numerical simulation can provide the required gas temperature distributions. This method will be implemented in our forthcoming investigations.

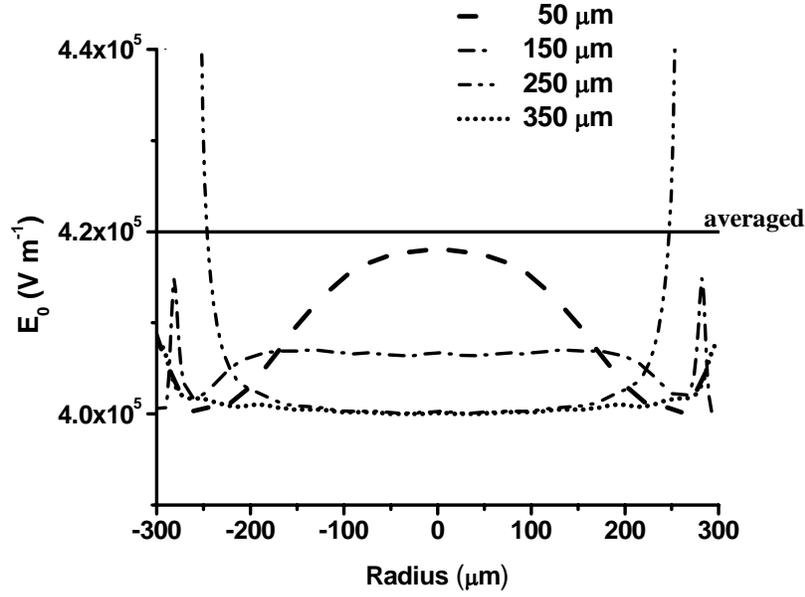

**Figure 6.** Space-resolved HF electric field (E$_0$ in V·m$^{-1}$) at distances 50 µm, 150 µm, 250 µm and 350 µm from the tip of the copper wire along the diameter of the active microplasma volume. Solid line presents electric field "averaged" value $E_{0.}$ = 4.2·10$^5$ V·m$^{-1}$ measured using OES.

Using the 3-D space-resolved absolute intensities obtained after inverse Abel transformation, the corresponding k$_{N_2(C)}$ values determined using space-resolved electric field values, density of nitrogen molecule at gas temperature, and plasma volume of each pixel space-resolved electron density is determined (figure 7). The averaged electron density determined using (11) taking into account averaged k$_{N_2(C)}$ and averaged I$_{N_2(C-B,0-0)}$ is 7.0·10$^{20}$ m$^{-3}$. The spatial distribution of electron density in figure 7 is similar to that determined by simulation [15].

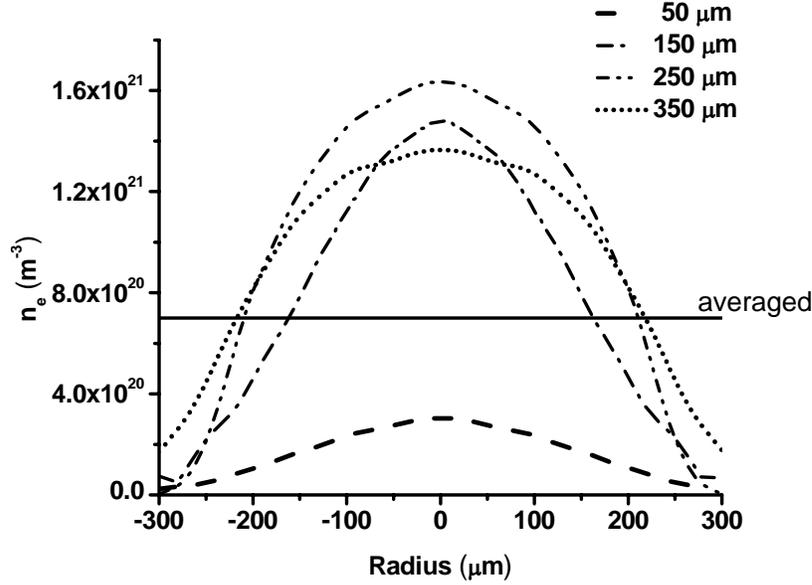

**Figure 7.** Space-resolved electron density ($n_e$ in m$^{-3}$) at distances 50 µm, 150 µm, 2540µm and 350 µm from the copper wire along the diameter of the observed active plasma volume. Averaged $n_e$ = 7.0·10$^{20}$ m$^{-3}$ determined applying OES with low space-resolution.

Applying determined plasma parameters production of chemical active species namely nitrogen atoms is determined with space resolution. The averaged production of nitrogen atoms is determined using averaged parameters amounts to about 9.6·10$^{26}$ m$^{-3}$·s$^{-1}$.

We determine plasma parameters applying space resolved OES and numerical simulation in local approximation with spaced resolution of 1.2 µm. Reliability of this resolution must be tested because local approximation means acceleration and loss of kinetic energy of electrons in the same point of space. To test validation of local approximation at different plasma conditions, plasma simulation at variable gas mixtures, pressure, frequency and space structure of electric field can be applied. Simplest test of validity of local approximation at high resolution is comparison of electron energy relaxation length $\lambda_\varepsilon = (\lambda \cdot \lambda^*)^{1/2}$ [16] with space resolution, where $\lambda$ and $\lambda^*$ are mean free path for elastic and inelastic collisions of electrons, correspondingly. Energy relaxation length of electrons at our experimental conditions calculated applying cross sections [10] amounts to 0.6 µm that is two times lower than applied space resolution.

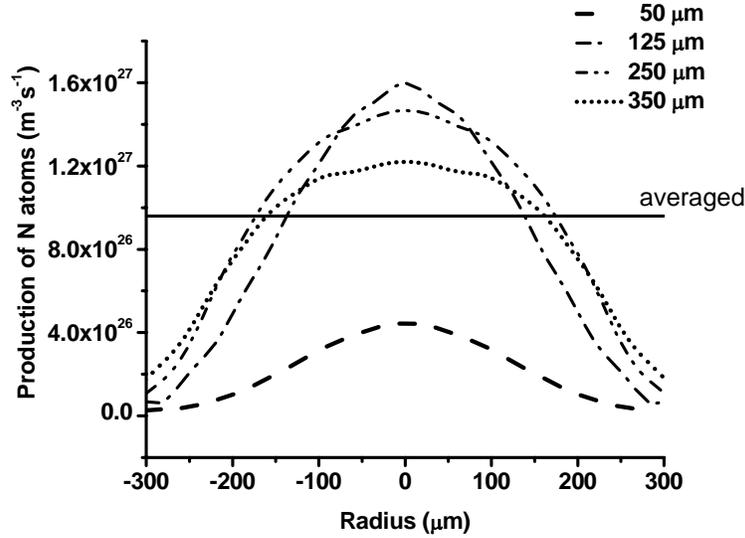

**Figure 8.** Space-resolved production of nitrogen atoms (m$^{-3}\cdot$s$^{-1}$) due to electron-impact dissociation of nitrogen molecules, at distances 50 µm, 150 µm, 250 µm and 350 µm from the copper wire along the diameter of the observed active plasma volume. Averaged production of nitrogen atoms (= 9.6·10$^{26}$ m$^{-3}\cdot$s$^{-1}$) calculated using the averaged electron density (in m$^{-3}$), averaged rate constant for electron-impact dissociation of nitrogen (m$^3\cdot$s$^{-1}$), and density of nitrogen molecules (m$^{-3}$) at gas temperature.

Comparison of plasma parameters determined with averaged OES and with space resolution (see figures 6,7) shows that these diagnostic methods provide consistent results and can be used for calculation of production of chemical active species (see figure 8) and simulation of chemical kinetics.

## 6. Conclusion

Space-resolved plasma characterization is achieved using OES, numerical simulation and microphotography in HF nitrogen plasma. Both direct and step-wise excitations of nitrogen emission are considered. Photoemissions imaged by a CCD camera are calibrated using the same emissions measured by OES. Space-resolved absolute intensities of nitrogen emissions are used for the determination of space-resolved plasma parameters such as electric field and electron density. Space-resolved rate constants for electron-impact dissociation of nitrogen molecules are determined, and used to estimate the production of nitrogen atoms with space resolution. It is observed that the space-resolved electric field, rate constant for dissociation of nitrogen molecule, electron density and production of nitrogen atoms are in good agreement with the corresponding space-averaged values.

## Acknowledgement

Financial support from the German Research Foundation (DFG) within the framework of project C2 of 'FOR1123-Physics of Microplasmas' is gratefully acknowledged.